\documentclass[12pt,reqno]{amsart}
\newcommand{\ver}{{\it }}

\usepackage{latexsym}
\newcommand{\R}{{\mathbb R}}
\newcommand{\Z}{{\mathbb Z}}
\newcommand{\C}{{\mathbb C}}

\newtheorem{theo}{{\sc Theorem}}[section]

\newtheorem{conj}[theo]{{\sc Conjecture}}

\newtheorem{lem}[theo]{{\sc Lemma}}
\newtheorem{prop}[theo]{{\sc Proposition}}

\newenvironment{rem}{\medskip\noindent{\it Remark:\/} }{\medskip}

\title[Macdonald's identities and the large $N$ limit\\ of $YM_2$ on the cylinder]
{Macdonald's identities and the large $N$ limit \\  of $YM_2$ on
the cylinder  }

\author{ Steve Zelditch \ver}
\address{Department of Mathematics, Johns Hopkins University, Baltimore,
MD
21218, USA}
\email{zelditch@math.jhu.edu}

\thanks{Research partially supported by NSF grant \#DMS-9703775 .}

\date{\today}

\begin{document}

\maketitle

\begin{abstract} The purpose of this paper is to
 determine the large $N$ asymptotics of the free energy
$F_N(a , U | A)$ of $YM_2$ (two-dimensional Yang Mills theory)
with gauge group $G_N = SU(N)$ on a cylinder where $a$ is a
so-called principal element of type $\rho$. Mathematically,
$$F_N(U_1, U_2 | A) = \frac{1}{N^2} \log H_{G_N}(A/2N,U_1, U_2) $$
is the central heat kernel of $G_N$. We find that $$F_N(a_N, U_N |
A) \sim \; \frac{N}{A} \;\; \Xi(d\theta, d \sigma)$$ where $\Xi$
is an explicit quadratic functional in the limit distribution $d
\sigma$ of eigenvalues of $U_N$,  which depends only on the
integral geometry of $SU(2)$. The factor of $N$ appears to
contradict some predictions in the physics literature on the large
$N$ limit of $YM_2$ on the cylinder (due to Gross-Matytsin,
Kazakov-Wynter others).
\end{abstract}

\section{Introduction} The purpose of this note is to determine
the large $N$ asymptotics of many   cases of the free energy
$F_N(U_{C_1}, U_{C_2} | A)$ of $YM_2$ (two-dimensional Yang Mills
theory) with gauge group $SU(N)$ on a cylinder.  Interest in this
large $N$ limit problem was raised around ten years ago in a
series  of physics papers by Douglas, Kazakov, Wynter, Gross,
Matytsin and others \cite{DK, KW, GM1, GM2}. They predicted,  on
the basis of physical and formal mathematical arguments, that the
large $N$ limit of $F_N(U_{C_1}, U_{C_2} | A)$ should be
well-defined and  related to the action along a path of the
complex Burgers equation with boundary densities determined by the
limiting eigenvalue densities of $U_{C_1}, U_{C_2}$. In certain
cases they predict a phase transition between a weak coupling and
a strong coupling regime.

Our results give the  asymptotics of $F_N(a_N ,U_N | A)$ where
$a_N$ is a {\it principal element of type $\rho$} of $SU(N)$ in
the sense of \cite{Ko} (see Section \ref{MACDO}), and where the
second sequence $\{U_N\}$ can be any sequence of elements of
$SU(N)$ which possesses limiting eigenvalue density $d\sigma$. The
limiting eigenvalue density of $a_N$ is uniform measure $d \theta$
on $S^1$. Our asymptotics reveal some surprising results:
\begin{itemize}

\item   $F_N(U_{C_1}, U_{C_2} | A) \sim \frac{N}{A}\; \;
\Xi(d\theta, d \sigma)$ for a certain (explicit) functional $\Xi$,
unlike the predictions that it tend to a limit functional as $N
\to \infty$. The extra factor of $N$  signals an error in some of
the heuristic physics arguments, and (as emphasized to the author
by M. Douglas and C. Vafa)  calls into question whether the large
$N$ of $YM_2$ on the cylinder is actually a string theory as
discussed in \cite{GM1, GM2, KW} and elsewhere. At least, it
indicates that extra hypotheses on the matrix pairs $U_{C_1},
U_{C_2}$ are necessary for the conjectured picture to be correct.

\item The principal asymptotic term is visibly analytic in
$A$, i.e. it never exhibits a phase transition to leading order,
even though $d \sigma$ could be any probability measure on $S^1$.
This aspect of the results is  consistent with the physics
predictions, in that the boundary condition  $d\theta$ puts the
system  in its strong coupling regime and it should therefore not
exhibit a phase transition between weak and strong coupling
\cite{D}.

\end{itemize}

The calculation presented here appears to be the first rigorous
calculation of the large $N$ limit of the partition function for
$YM_2$ on the cylinder. Hence, it is unclear at this time whether
the anomalous factor of $N$ only occurs in the special case where
$U_{C_1} = a_N$ is a principal element of type $\rho$ or whether
it holds more generally in this context.

 We
should emphasize that the other rigorous mathematical results (of
which we are aware) have largely confirmed the large $N$ limit
picture developed in \cite{DK, GM1, GM2, KW} and elsewhere. The
Douglas-Kazakov phase transition of the genus $0$ partition
function at $A = \pi^2$ has been proved by Boutet de Monvel -
Shcherbina in \cite{BS}. A large deviations analysis of spherical
integrals by Guionnet-Zeitouni in \cite{GZ} justifies some of the
predictions by Matytsin \cite{M} and Gross-Matytsin \cite{GM1} on
the asymptotics of characters $\chi_R(U)$. However, the results of
\cite{GZ} pertain to the analytic continuation of $\chi_R$ to
positive elements of $GL(N, \C)$ and it is possible that
Matytsin's predictions fail when $U = a_N$. We intend to explore
this question in a future article.

To state our results, we introduce some notation.
 The  partition function of  $YM_2$  on a cylinder, with gauge group
  equal to  $G$, is given by Migdal's formula (see \cite{W, W2}):  \begin{equation} \label{KERNEL} {\mathcal
  Z}_G
(U_1, U_2 | A) = \sum_{R \in \widehat{G}} \chi_{R}(U_1)
\chi_R(U_2^*) e^{- \frac{A}{2N} C_2(R)} \end{equation} Here, $A
\geq 0$ is  the area of the cylinder, and
 the sum  runs over the irreps (irreducible
representations) of $G$, with $\chi_R$ the character of $R$ and
with $C_2(R)$ equal to the eigenvalue of the Casimir  $\Delta$ in
the irrep $R$.  Thus, ${\mathcal Z}_G $ is the value at time $t =
\frac{A}{2N}$ of the  {\it central } heat kernel of $G$:
 \begin{equation} \label{CCKERNEL} H_{G}(t,U_1, U_2) = \sum_{R \in \widehat{G}} \chi_{R}(U_1)
\chi_R(U_2^*) e^{- t C_2(R)},  \end{equation} i.e.   the kernel of
the heat operator acting on the space of central functions on
$SU(N)$. It is obtained from the  usual heat kernel by averaging
both variables over conjugacy classes. Since ${\mathcal Z}_G (U_1,
U_2| A) $ is conjugacy invariant, one may assume that $U_1, U_2$
are diagonal and we write $U_j = D(e^{i \theta_1^j}, \dots, e^{i
\theta_{\ell}^j})$, where $\ell $  is the rank of $G$ (i.e. the
dimension of its maximal torus). The main quantity of interest is
the  free energy, defined  by
\begin{equation}F_G(U_{C_1}, U_{C_2} | A) = \frac{1}{N^2} \ln {\mathcal Z}_G (U_{C_1}, U_{C_2} | A).
\end{equation}

We now consider the large $N$ limit of the partition function and
free energy. The large $N$ limit refers to an increasing  sequence
$G_N $ of groups, e.g. the classical groups $G_N = U(N), SU(N),$ $
SO(N), Spin(N).$ For the sake of simplicity we restrict attention
to $SU(N)$ and we abbreviate $F_{SU(N)}$ by $F_N$ (etc.) The limit
we are interested in is a pointwise limit of the central heat
kernel, which obviously requires some discussion since the space
on which the heat kernel is defined changes with $N$.

 The large $N$ limit of $F_N$ is defined as follows: take a pair
of sequences $\{U_{N j}\}, U_{N j} \in SU(N)$ ($j = 1,2$) of
elements whose eigenvalue distributions $$d\sigma_{N j} : =
\frac{1}{\ell_N} \sum_{k = 1}^{\ell_N} \delta (e^{i \theta_k^{N
j}}) \;\; \in \; {\mathcal M}(S^1)$$  tend to a limit measures
$\sigma_j$. Here,  $\ell_N = N - 1$ denotes the rank of the group,
and  ${\mathcal M}(S^1)$ denotes the probability measures on the
unit circle, and convergence is in the weak sense of measures. We
will denote this situation by $U_{N j} \to \sigma_j$. That is,
\begin{equation} U_{N j} \to
 \sigma_j \iff \frac{1}{\ell_N} \sum_{k = 1}^{\ell_N} \delta (e^{i \theta_k^{N j}}) \to \sigma_j
  \in {\mathcal M}(S^1),\;\; (j = 1, 2). \end{equation}

\begin{conj} \label{LARGEN} \cite{GM1} (pages 8-9) Assume that $U_1 \to \sigma_1, U_2 \to \sigma_2$. Then
 $$\begin{array}{l} F_N(U_1, U_2 | A) \to F(\sigma_1(\theta), \sigma_2(\theta)| A) = S(\sigma_1(\theta), \sigma_2(\theta)| A) \\  \\  -
\frac{1}{2} \int_{S^1} \int_{S^1} \sigma_1(\theta) \sigma_1(\phi)
\log |\sin \frac{\theta - \phi}{2}| d\theta d \phi   -
\frac{1}{2} \int_{S^1} \int_{S^1} \sigma_2(\theta) \sigma_2(\phi)
\log |\sin \frac{\theta - \phi}{2}| d\theta d \phi, \end{array}$$
where the functional $S$ is a solution of the Hamilton Jacobi equation
$$\frac{\partial S}{\partial A} = \frac{1}{2} \int_0^{2\pi} \sigma_1(\theta)
[(\frac{\partial}{\partial \theta} \frac{\delta S}{\delta \sigma_1(\theta)})^2
- \frac{\pi^2}{3} \sigma_1^2(\theta)]. $$ \end{conj}

It is often  assumed in the physics articles  that the limit
measures have densities. The function $S$ can be identified as the
action along the solution of the boundary value problem for
Hopf-Burgers equation
$$\left\{ \begin{array}{l} \frac{\partial f}{\partial t} + f \frac{\partial f}{\partial x}
= 0, \\ \\ \Im f(0, x ) = \sigma_1,\;\; \Im f(A, x) = \sigma_2.
\end{array}\right.$$

Under certain assumptions on the limit densities, it is further
conjectured that a third order phase transition between the weak
and strong coupling regimes  should occur. In the case of the disc
(the cylinder with $U_1 = I$), Kazakov-Wynter and Gross-Matytsin
have predicted a phase transition point at $$ A = \frac{\pi}{2
\Theta(\pi)}, \;\;\; \Theta(\theta) : = \int \frac{d
\sigma(\theta')}{\theta - \theta'}, $$ Here, $d \sigma$ is the
limit distribution for $U_2$ and the integral is presumably to be
understood as the Hilbert transform on $S^1$ (although it is
written as the Hilbert transform on $\R$). The weak coupling
regime is characterized by a gap in the support of the limit
density (i.e. an interval of $S^1$ on which $d\sigma_1$ vanishes).

The mathematical evidence for these  conjectures is largely based
on
 approximating  the  discrete sum (\ref{KERNEL})  over $R \in \hat{G}$ by a continuous (and
only partially defined) integral over a space of ``densities of
Young tableaux", to which the saddle point  method is applied (cf.
\S 3). These arguments are not rigorous and also make some
implicit assumptions (see the discussion after Theorem
\ref{RHON}).

The main observation of the present article is that there exists a
 simple, rigorous alternative for calculating the large $N$ limit
 when one of the arguments is a so-called principal element of type $\rho$.
 This  alternative method is based on the use  MacDonald's identities (in Kostant's formulation) to factor the  partition function
 as a product over positive roots. Background on MacDonald's
 identities will be provided in \S \ref{MACDO} (see also   \cite{MAC, Ko, F,
 PS}).

 The first result one obtains this way is a
 limit formula in which the  time variable in the
 partition function is {\it not} rescaled. We use
 the notation  $ d\sigma * \bar{d\sigma}(x)$ for the measure
 defined by
 \begin{equation} \label{STAR} \int_{S^1} f(e^{i x})  d\sigma *
 \bar{d\sigma}(x):= \int_{S^1} \int_{S^1} f(e^{i (x - y)})
 d\sigma(x) d\sigma(y). \end{equation}
Thus,  $*$ denotes convolution of measures and $d \bar{\sigma}$ is
short for $d \sigma(\bar{x})$. For notational simplicity we
sometimes denote $e^{ix} \in S^1$ more simply by $x$.

\begin{theo} \label{RHO} Let $k_N \in {\bf su}(N)$ be  diagonal matrices with entries
$\theta_j^N$, and assume $d\sigma_N:= \frac{1}{\ell_N} \sum_{j =
1}^{\ell_N} \delta(e^{i \theta^N_j}) \to \sigma.$ Then, as $N \to
\infty$,
$$\begin{array}{lll} \frac{1}{N^2} \log H_{SU(N)}(t,a_N, e^{k_N}) & \to  &  - \frac{1}{2} \;\log \eta (i t)
+\frac{1}{2}  \int_{\R} \log H_{SU(2)} ( t, e^{i x}, e^{\frac{i
\pi}{2}}) d\sigma * \bar{d\sigma}(x).
$$ \end{array}$$
\end{theo}

Here,  $H_{SU(2)}$ is the central heat kernel of $SU(2)$ and $e^{i
x}$ is short for the diagonal matrix $D(e^{i x}, e^{-i x})$. The
element $a$ is  the principal element of type $\rho$ of $SU(2)$,
namely $a = D(e^{\frac{i \pi}{2}}, e^{-\frac{i \pi}{2}})$.

 As discussed in \cite{F} (Proposition 1.3),
the central heat kernel of $SU(2)$ at these special values is
given by
\begin{equation} \label{HSU2} H_{SU(2)}( t, e^{i x}, e^{\frac{i \pi}{2}}) =
\frac{ \theta_1( e^{i \pi x}, i t)}{2 e^{- \pi t/4} \sin \pi x}
\end{equation}
where  \begin{equation} \label{THETA} \theta_1(z, t) = 2 \sum_{n =
0}^{\infty} (-1)^n \sin\{(2n + 1) z\} e^{- \pi (n + 1/2)^2 t}
\end{equation}  is Jacobi's theta function. MacDonald's identities
are more often stated in terms of Jacobi theta functions, but we
find it easier to work directly with heat kernels. We note that
the factor of $\pi$ in the exponent $ \pi (n + 1/2)^2 t$ is
responsible for the appearance of $\pi t$ in many expressions to
follow in the heat kernel.

At first sight, this result seems  to  explain the normalization
$\frac{1}{N^2} \log {\mathcal Z}_N$. However,  the large $N$ limit
conjectures concern the scaling limit of the central heat kernel
under  $t \to \frac{A}{2N}$. This puts into play a simultaneous
limit process in $d \sigma_N * \bar{d\sigma}_N \to d \sigma
* \bar{d\sigma}$ and in the asymptotics of theta functions.
We find that for our cases of the problem, this rescaling changes
the growth rate of the free energy.

\begin{theo} \label{RHON} Suppose that $e^{k_N}$ is a sequence such that  $d\sigma_N \to
d\sigma$. Then,
$$\begin{array}{lll} F_N(a_N, e^{k_N} | \frac{A}{2N})
& = & \frac{1}{2} \int_{S^1}  \log  H_{SU(2)}( A/2 N, e^{i x},
e^{\frac{i \pi}{2}}) d \sigma_N
* \overline{d \sigma_N}(e^{i x}) - \frac{1}{2} \log \eta(\frac{iA}{2N}) + O(1/N)\\ & & \\ &  \sim &
- \frac{N}{ A} \; \{  \int_{S^1  }  \frac{1}{\pi} \;\min
\{d(e^{ix}, e^{i \pi/2}), d(e^{ix}, e^{- i \pi/2})\}^2 d \sigma
* \overline{d \sigma}(e^{i x}) - \frac{\pi }{12 }\}, \end{array}$$
where $d(e^{ix}, e^{iy})$ is the distance along $S^1$.

\end{theo}

The minimal distance above arises as   the distance $d(C(a),
e^{ix})$ in $SU(2) $ of the diagonal element $D(e^{ix}, e^{-ix })$
to the conjugacy class $C(a)$ of $a$ of the principal element of
type $\rho$, which is conjugate to $D(e^{i \pi/2}, e^{-i \pi/2})$.

 We note that there are two terms of opposite sign in the leading
order term. We note that the terms cancel when $U_1 = a = U_2$,
since then $d \sigma * \overline{d\sigma} = \frac{d\theta}{2 \pi}$
and the first term reduces to $- \frac{N}{ A} \{\frac{1}{\pi}
\frac{4}{2 \pi} \frac{1}{3} (\frac{\pi}{2})^3 - \frac{\pi}{12}\} =
0.$ This appears to be the kind of case studied in \cite{GM1}. But
the two terms  cannot cancel in all cases, and indeed,  the
leading term does not not cancel in the simplest case, where $k_N
= 0$ for all $N$, i.e. where $U_2 = Id$. We then have (see
\cite{F1} or \cite{F}, Proposition 1.2)
 \begin{equation} \label{RHOI} H_{SU(N)} (t, a_N , I) = e^{-  \dim SU(N)\;  t/24}
 \eta(i t)^{\dim SU(N)},
 \end{equation}
where  $\eta(t)$ is Dedekind's $\eta$-function.  As is easy to see
(and will be verified below), the eigenvalue distribution  of
$a_N$ tends to $\frac{d \theta}{2 \pi}$, while that of $I$ is
obviously $\delta_1$. In this case, the asymptotic mass equals $1$
and the continuous term equals zero. We separate out this special
case since the result is most easily checked on this example:

\begin{prop}\label{ID}  When $U_1 = a_N$ and $U_2 = I$, so that
$\sigma_1 = d \theta, \sigma_2 = \delta_1$, then
$$\begin{array}{lll} \frac{1}{N^2} \log   {\mathcal Z}_{SU(N)} (e^{4 \pi
i \rho}, 1 | A) = -\frac{2N}{ A} \{\frac{\pi  }{12  }\} -
\frac{1}{2} \log (\frac{A}{2N}) - A/48 N + O(e^{- c N}).
\end{array}
$$

\end{prop}

Note that  Theorem \ref{RHON} reduces to Proposition \ref{ID} when
$d \sigma = \delta_0$.

Some final comments on the anomalous extra factor of $N$. The
prediction that $\frac{1}{N^2} \log {\mathcal Z}_N (A)$ should
have  a limit determined by a variational problem is one of many
predictions of this kind in field theory and statistical mechanics
and it seems very strange that an extra factor of $N$ should
appear in our calculations.
  It is a mystery how it would appear in
the graphical or diagrammatic calculations which initially
suggested that the large $N$ limit of gauge theory is a string
theory, i.e. without using (\ref{KERNEL}).  It seems best to leave
it to the string theorists to decide how much the anomalous factor
of $N$ affects the picture of the large $N$ limit of gauge theory
as a string theory.

We are on somewhat safer ground in trying to account for the extra
factor of $N$ in the formal calculations based on (\ref{KERNEL}).
 From discussions with M. Douglas and V. Kazakov, it seems
plausible that the anomaly is due to the unusual behavior of the
character values $\chi_R(a)$ at the principal element of type
$\rho$. It was proved by Kostant \cite{Ko} that the only character
values at this element are $\chi_R(a) = -1, 0, 1$ and it is
reasonable to expect that the value oscillates regularly between
these values. It appears that the physics predictions implicitly
assumed a less oscillatory behaviour in character values, and in
particular less oscillation in the signs of character values. The
heuristic calculations in \cite{GM1, GM2, KW} of the large $N$
limit of the free energy on the cylinder were based on special
cases (such as $U_2 = U_1^*$) which do not have such oscillations
in sign. The sign oscillation causes much more cancellation than
expected, and this could explain why our asymptotics have the form
$e^{- N^3 \Xi}$ rather than $e^{- N^2 \Xi}$. These special values
$\chi_R(a)$ might also be inconsistent with Matytsin's character
asymptotics, and we plan to check this in the future.

This paper is organized as follows.  In Section \ref{MACDO}, we
review MacDonald's identity and associated objects which we will
use in the proof of the main result. The main results are proved
in Section \ref{PROOF}.  In Section \ref{FR}, we discuss some
aspects of the proof and mention some other results which could be
proved by the methods of this paper.
\medskip

\noindent {\bf Acknowledgements} The author would like to thank M.
Douglas, P. Etingof,  V. Kazakov, B. Kostant, N. Reshtikhin, T.
Tate,  and C. Vafa  for going over some of the details of the
calculations and for comments (quoted above) on the relations
between our results and the predictions in the physics papers.
\medskip

\section{\label{MACDO} Background on MacDonald's identities} In this section, we   review the $\eta$ function and MacDonald's
identity. Our main references are the articles \cite{Fr} of I.
Frenkel and those  \cite{F, F1} of H. Fegan. Since the  articles
employ very different, possibly confusing,  notational
conventions, we highlight the main ones:
\begin{itemize}

\item In \cite{Fr}, the $\eta$-function is defined in a
non-standard way in the lower half-plane (see \S 4.4) and
consequently the heat kernel is evaluated at time  $t = 4 \pi i b$
where $\Im b < 0$. We will quote results of \cite{Fr} in terms of
$t$ and we will use the usual definition of $\eta$ as a modular
form on the upper half plane.

\item In \cite{Fr}, the special element of type $\rho$ is denoted
$e^{4 \pi i \rho}$, but the exponent refers to the dual element of
the Cartan subalgebra rather than the linear functional $\rho$. To
avoid confusion, we denote it simply by $a$ as in \cite{F}.

\item Fegan works with the Schrodinger equation rather \cite{F1}
(1.1) rather than the heat equation, so our formulae differ from
his in that his $it$ is our $t$.

\end{itemize}

\subsection{$\eta$ function}

The   Dedekind eta-function \cite{I, A} is defined by
\begin{equation} \label{eta} \eta(z) = e^{2  \pi i z/24} \;
\Pi_{n=1}^{\infty} (1 - e^{ 2 \pi i  n z}), \;\;\;\; \Im z > 0.
\end{equation}  It is a modular cusp form of weight $1/2$, i.e. it
satisfies
$$\eta (
\gamma z) = \theta (\gamma) j_{\gamma}(z)^{1/2}
\eta(z),\;\;\; \mbox{if}\;\; \gamma \in SL(2, \Z), $$ where
$\theta(\gamma)$ is a certain multiplier which we will not need to
know in detail (see \cite{I}, \S 2.8). When $\gamma z = - 1/z$, we
have
$$\eta(-\frac{1}{z}) = (i z)^{1/2} \eta(z).$$

In studying the free energy, we are particularly interested in
$\log \eta (i y)$ as $y \to 0 +$. The transformation law for $\log
\eta (i y)$ goes as follows \cite{A}:  For real numbers $y
> 0$ we have:
\begin{equation} \label{LOGETA}\begin{array}{lll} \log \eta(i y) & =
& -\frac{\pi}{12 y} - \frac{1}{2} \log y + \sum_{m = 1}^{\infty}
\frac{1}{m} \frac{1}{1 - e^{2 \pi m / y}}. \end{array}
\end{equation}

\subsubsection{Jacobi's theta function}

Although we mainly use the heat kernel parametrix on $SU(2)$ to
obtain asymptotic formulae, the alternative in terms of Jacobi's
theta function (\ref{HSU2}) can be used to check the details. The
small time expansion of the heat kernel can be obtained form
Jacobi's imaginary  transformation law:
$$\theta_1( -x/ \tau, - 1/ \tau) =  i (- i \tau)^{1/2} e^{- i x^2/ \tau} \theta(x, \tau). $$
 Thus, we have:
\begin{equation} \label{JTL} \theta_1(x, \frac{i A}{2 N})  =
i ( \frac{ 2 N}{A})^{1/2} e^{-  2 x^2 N / A \pi} \;
\theta_1(\frac{2 i N}{ A} x, \frac{2i  N}{ A})
\end{equation}
 Here, we use that
$$\tau = \frac{i A}{2 N} \implies -1/ \tau  = i
\frac{2 N}{A} \implies e^{ -\pi \frac{ A}{2 N}} \to e^{- \pi
\frac{2 N}{A}}. $$ The large $N$ or small time asymptotics now
follow from (\ref{THETA}).

\subsection{Central heat kernels}

Let $G$ be any compact, connected Lie group.  We denote by $R$ the
root system of $({\bf g}_{\C}, {\bf h}_{\C})$,  where $ {\bf h}$
is its Cartan subalgebra.  We denote by $R_+$ the positive roots,
by $P$ the lattice of weights and by $P_{++} \subset P$ the
dominant weights.

We recall that the eigenvalue of the Casimir operator
(bi-invariant Laplacian) $\Delta$ of $G$ in the representation
with highest weight $\lambda$ is equal to
$$\Delta|_{V_{\lambda}} = (||\lambda + \rho||^2 - ||\rho||^2) Id |_{V_{\lambda}},\;\;\;
\rho = 1/2 \sum_{\alpha \in R_+} \alpha. $$ The fundamental
solution of the heat equation is given by $k_(t, x, y) = \nu(x
y^{-1}, t)$ where
\begin{equation} \nu(g, t) =  \sum_{\lambda \in P_{++}} (\dim
V_{\lambda} ) \;  \chi_{\lambda}(g) e^{- t/2 (||\lambda + \rho||^2
- ||\rho||^2)},
\end{equation}
As in \cite{Fr}, \S 4.3 it then follows that
 the central heat kernel is given by
\begin{equation} \label{CHK} H_{G} (t, h, k) =  \sum_{\lambda \in P_{++}} \chi_{\lambda}(e^{h})
\chi_{\lambda}(e^{-k}) e^{- t/2 ( ||\lambda + \rho||^2 -
||\rho||^2)}.
\end{equation}
This is simply a different notation for (\ref{CCKERNEL}).

\subsection{$SU(2)$}

As was recognized clearly by H. Fegan and others, the central heat
kernel of $SU(2)$ plays an important role in MacDonald's
identities.  There is one positive root $\alpha$, which can be
identified with $1$ if we choose it as the basis of the Cartan
dual subalgebra. Then $\rho = \frac{1}{2}$ and the Killing form is
$B(x, y) = \frac{1}{2} xy.$ In this case, the principal element
$a$ of type $\rho$ has eigenvalues $ e^{\pm \frac{i \pi}{2}}$. The
weight lattice is $\frac{1}{2} \Z$. The character of the
irreducible representation of highest weight $\lambda$ is
$\chi_{\lambda}(x) = \frac{\sin (2 \lambda + 1) x}{\sin \pi x}.$
We have:
$$\chi_{\lambda}(a) = \left\{\begin{array}{ll}  -1, & \lambda \;\;
\mbox{is an odd integer} \\ & \\
0, & \mbox{is not an  integer} \\ & \\
1, & \mbox{is an even integer} \end{array} \right. $$ The
eigenvalues of the Casimir are $c(\lambda) = \frac{1}{2} \lambda
(\lambda + 1)$.

The central heat kernel $H(t, a, y)$ at the special point $a$ may
be expressed in terms of Jacobi's theta function as in
(\ref{HSU2}).  As a special case of MacDonald s identities, we
further have (cf. \cite{F1}, (3.12)):
\begin{equation} \label{HETA} H_{SU(2)}(t, a, 1) = (e^{- \pi t/12}
\eta(i t))^3. \end{equation}

\subsection{MacDonald identities}

We briefly review MacDonald's identities for   a compact,
semi-simple, simply connected Lie group $G$  and its relation
(proved by Kostant \cite{Ko})  to the central heat kernel. They
are thus valid for $SU(N)$.  We follow \cite{F, F1, Fr}.

MacDonald's identity involves a conjugacy class $C_a$ of special
elements of $G$, which we will denote by $a$,  which are conjugate
to $e^{2 x_{\rho}}$ where $x_{\rho}$ is dual under the Killing
form to $\rho$.  Such elements are called `principal elements of
type $\rho$'.  A principal element of type $\rho$ is characterized
in \cite{Ko} (see 1.3 p. 181) as a regular element such that the
order of $Ad(a)$ equals $h$ (the order of the Coxeter element).
One has $h = N$ for $SU(N)$, so the eigenvalues of $Ad(a)$ must be
distinct $N$th roots of unity. Hence
\begin{equation} a_N \to \frac{d\theta}{2 \pi}, \end{equation}
i.e. the eigenvalues of $a_N$ become uniformly distributed in the
large $N$ limit. A detailed description of such elements  can be
found in \cite{Ko, F1, AF}.

 In  Kostant's formulation, as described
by Frenkel,  the MacDonald's identities take the form (\cite{Fr},
Proposition (4.4.5))
\begin{equation} \label{KOSTANT} \theta(k, 4 \pi i t) = \sigma(k) \sum_{\lambda \in P_{++}}
 \chi_{\lambda}(a)
\chi_{\lambda}(e^{-k}) e^{- t/2 ||\lambda + \rho||^2},
\end{equation}
where (see \cite{Fr}, (1.1.8)
$$\sigma(k) = \Pi_{\alpha \in R_+} (e^{\langle\alpha, k \rangle /2} -
e^{-\langle  \alpha, k \rangle /2}),$$ where $\theta(k, i t)$ is
the theta-function defined in Definition (4.4.1) of \cite{Fr}: For
$t >  0$,
\begin{equation}\label{FRANKEL}  \begin{array}{lll} \theta(k, i t) & = & e^{- 2 \pi ||\rho||^2}
\Pi_{\alpha \in R_+} (e^{\frac{\langle \alpha, k \rangle}{2}} -
e^{- \frac{\langle \alpha, k \rangle}{2}}) \\ & & \\ & \cdot &
\Pi_{n=1}^{\infty}(1 - e^{- 2 \pi  n t})^{\ell} \Pi_{\alpha \in R}
(1 - e^{- 2 \pi  n t+ \langle \alpha, k \rangle})
\end{array} \end{equation}
 For our purposes the key formula is the following (\cite{Fr}, Proposition
 (4.4.4)):
\begin{equation} \label{ROOTPROD}  \theta(k, it ) = \eta(i t)^{- |R_+| + \ell} \Pi_{\alpha \in R_+}
\theta_1 (\langle \alpha, k \rangle), i t ). \end{equation}

We put  $\tilde{\theta} (x , it ) = \frac{\theta_1(x , i t)}{2
e^{- t/4}  \sin \pi x}.$ The following is the version of
MacDonald's identities proved in  Theorem 1.5 of \cite{F1}
(together with its Proposition 1.3). As above, an element $e^{ i
\theta}$ is identified with a diagonal element $D(e^{i \theta},
e^{- i \theta})$ of $SU(2)$.

\begin{lem} \label{MCDHEAT} Let $G$ be compact, connected, semi-simple and simply
connected, and let $a$ be an element of type $\rho$. Then $$ H_{G}
(t, a, e^{-k}) = (e^{- \pi t/12} \eta( i t))^{- |R_+| + \ell}
\Pi_{\alpha \in R_+} H_{SU(2)} (t, e^{i \langle \alpha, k
\rangle}, e^{i \pi/2}).$$
\end{lem}

\begin{proof}

Puttting together (\ref{KOSTANT}), (\ref{ROOTPROD}) and
(\ref{CHK}), we get
\begin{equation} \label{RHOFORM}\begin{array}{lll}  H_{G} (t, a, e^{-k})
&  = &  \sigma(k)^{-1}  e^{ A/ 4 N ||\rho||^2} \theta(k, i t) \\ & & \\
&  = &  e^{ t ||\rho||^2} \eta( i t )^{- |R_+| + \ell} \Pi_{\alpha
\in R_+} \tilde{\theta} (\langle \alpha, k \rangle , i t
).\end{array}
\end{equation}
The stated result now follows from (\ref{HSU2}).
\end{proof}

As mentioned above, the simplest case (\ref{RHOI}) comes from
combining Lemma \ref{MCDHEAT} and (\ref{HETA}):
 \begin{equation} \label{RHOI} H_{G} (t, a , I) = e^{-  \dim G\;  t/24}
 \eta(i t)^{\dim G},
 \end{equation}

\section{\label{PROOF} Large $N$ limit of ${\mathcal Z}_{SU(N)}(a, U |A)$: Proof of Theorem \ref{RHON}}

\subsection{Simplest case}

We first consider the simplest case (\ref{RHOI}). Since $k = 0$
(i.e. $U_2 = I$), we are essentially dealing with $YM_2$ on the
disc. This case is discussed in detail in \cite{GM2}, \S 4.

\begin{prop} When $U_1 = a$ and $U_2 = 1$, so that
$\sigma_1 = d \theta, \sigma_2 = \delta_1$, then
$$\begin{array}{lll} \frac{1}{N^2} \log   {\mathcal Z}_{SU(N)} (a, 1;A) = -\frac{2 N }{ A } \frac{\pi}{24} - \frac{1}{2}
\log (\frac{A}{2N}) - A/48 N + O(e^{- c N}).
\end{array}
$$
Thus, no phase transition occurs.

\end{prop}

\begin{proof}

Macdonald's identity gives:
\begin{equation} \label{RHOFORM}\begin{array}{lll}  {\mathcal Z}_{SU(N)} (a, 1;
A) &  = & e^{-  \pi \dim SU(N)  A/24 N } \eta(i A/2N)^{\dim
SU(N)},\;\; .\end{array}
\end{equation}

The free energy is then
\begin{equation} \label{LOGRHOI}  \frac{1}{N^2} \log  H_{SU(N)} (A/2N , a , I)
= \frac{ \dim SU(N)}{N^2}\{-  \frac{ A \pi}{48 N} + \log  \eta(i
A/2 N)\}.
\end{equation}

We note that $\frac{ \dim SU(N)}{N^2} = \frac{1}{2} +
O(\frac{1}{N}).$
 We substitute $y =
A/2 N$ in the right side of (\ref{LOGETA}) to get:

\begin{equation} \log \eta(i A/ 2N) =
 -\frac{2N}{A} \frac{\pi  }{24 } - \frac{1}{2} \log (\frac{A}{2N})  + \sum_{m = 1}^{\infty}
\frac{1}{m} \frac{1}{1 - e^{4 N \pi m / A}} \end{equation}

The summand of the sum  $\sum_{m = 1}^{\infty} \frac{1}{m}
\frac{1}{1 -e^{4 N \pi m / A}}$  is smaller than $1/2 e^{- 4 N
(\pi/A) m} )$ for $N$ sufficiently large, so we may bound the sum
by the geometric series and obtain a bound  of
$$\frac{1}{1 - e^{- 4 N (\pi /A) } } - 1 \leq C e^{- 4 N
(\pi /A)} ,\;\; \mbox{for}\; N \; \mbox{sufficiently large}. $$
Thus, the sum is an exponentially small correction, and we have
\begin{equation} \label{RHOFORMA}\begin{array}{lll} \frac{1}{N^2} \log   {\mathcal Z}_{SU(N)} (a,
1;A) & = & \frac{ \dim SU(N)}{N^2}\{-  A/48 N   -\frac{\pi N }{6 A } - \frac{1}{2} \log (\frac{A}{2N}) \} + O(e^{-c N/A}). \\ & & \\
& \sim &  -\frac{\pi  }{24  }\frac{2N}{A} - \frac{1}{2} \log
(\frac{A}{2N}) - A/48 N.
\end{array}
\end{equation}

\end{proof}

We see clearly the anamolous factor of $N$. Also,  we see that
there is no phase transition, despite the fact that the two
eigenvalues densities, $\delta (1), d\theta$ have very different
support properties. In almost exactly this case, the lack of phase
transition is predicted by Gross-Matytsin in \cite{GM2}, \S 4.
They explain that when the support of $\sigma_1$ is the whole
circle and the support of $\sigma_2$ is a single point, then the
system is always in the strong coupling phase.

They further explain that when  $\sigma_1 = \delta(\theta = 0)$,
the solution of the Hopf equation can be obtained as the solution
of the integral equation
$$(1 - \frac{t}{A}) f(t, \theta) = \frac{- \theta}{A} + \int
\frac{\sigma_1(\theta') d \theta'}{\theta - \theta' - t f(t,
\theta)}. $$ In our case, $\sigma_1 = 1$ so the equation is simply
$$(1 - \frac{t}{A}) f(t, \theta) = \frac{- \theta}{A} - \log
(\theta -   t f(t, \theta)). $$ It would be interesting to solve
this equation and check that the solution is analytic in $A$.

\subsection{Proof of Theorem (\ref{RHO})}

We may take $e^k$ to be  a diagonal matrix with entries $e^{2 \pi
i \lambda_j(N)}$. First, we put $t = A/ 2N$. By  Lemma
\ref{MCDHEAT} we have:
\begin{equation}
\begin{array}{lll} \frac{1}{N^2} \log H_{SU(N)} (t, a, e^k) & = &
 \frac{1}{N^2} \{  (- |R_+| + N) [- \frac{\pi t}{12} +
  \log \eta (-i   t/4 \pi ) \}\\ & & \\ & + & \frac{1}{N^2} \sum_{\alpha > 0}
  \log  H_{SU(2)} (t,  e^{i \langle \alpha, k \rangle}, e^{i \pi/2} )
\end{array} \end{equation}
We note that both sides of this equation are real, so that we take
take the real part $\Re$ without changing the equation.

We now specialize to the case of $SU(N)$. Its roots are $e_i -
e_j$ and its positive roots satisfy $i < j$. Hence,  $\langle
\alpha, k \rangle = \lambda_i - \lambda_j.$  Hence
\begin{equation}\begin{array}{lll}   \sum_{\alpha \in R_+} \log  H_{SU(2)}
(t,  e^{i \langle \alpha, k \rangle}, e^{i \pi/2} )  & = & \sum_{i
< j} \log  H_{SU(2)} (t,  e^{i (\lambda_i - \lambda_j)},
 e^{i \pi/2})\\ & & \\
 & = & \frac{1}{2} \sum_{i \not= j} \log  H_{SU(2)} (t, e^{i (\lambda_i - \lambda_j)},
 e^{i \pi/2} ). \end{array} \end{equation}
In the last equality we use that $H_{SU(2)} (t, e^{i (\lambda_i -
\lambda_j)},
 e^{i \pi/2} ) = H_{SU(2)} (t, e^{i (\lambda_j - \lambda_i)},
 e^{i \pi/2} ). $

 We further have
$$ \sum_{i \not= j} \log  H_{SU(2)} (t, e^{i (\lambda_i - \lambda_j)},
 e^{i \pi/2} ) =  \sum_{i,  j} \log  H_{SU(2)} (t, e^{i (\lambda_i - \lambda_j)},
 e^{i \pi/2} ) - N \log H_{SU(2)}(t, 1, e^{i \pi/2}).$$

Since $\ell_N \sim N$ for $SU(N)$, we have
$$\begin{array}{lll} \frac{1}{N^2} \sum_{\alpha \in R_+} \log  H_{SU(2)}
(t,  e^{i \langle \alpha, k \rangle}, e^{i \pi/2} )
 & = & \int_{S^1} \int_{S^1}  \log  H_{SU(2)} (t, e^{i x}, e^{i \pi/2} )
d \sigma_N(e^{i (x- x')}) \overline{d \sigma_N}(e^{i x'})\\ & & \\ & &  - \frac{1}{N} \log H_{SU(2)}(t, 1, e^{i \pi/2})  \\ & & \\
& = & \int_{S^1}  \log  H_{SU(2)} (t, e^{i x}, e^{i \pi/2} ) d
\sigma_N * \overline{d \sigma_N}(e^{i x})  \\ & & \\ & &  -
\frac{1}{N} \log H_{SU(2)}(t, 1, e^{i \pi/2})
\end{array}$$ where
$$d \sigma_N * \overline{d \sigma_N}(e^{ix }) = \int_{S^1} d \sigma_N(e^{i (x - x') } ) \; d \sigma_N (e^{i x'}).$$

If  $$\sigma_N:= \frac{1}{\ell_N} \sum_{j = 1}^{\ell_N}
\delta(e^{2 \pi i \lambda_j(N)}) \to \sigma \in {\mathcal
M}(S^1),$$  then
$$d \sigma_N * \overline{d \sigma_N} \to d \sigma * \overline{d \sigma},$$
since the Fourier coefficients of the left side tend to those of
the right side. Thus, we obtain the stated limit.

\subsection{Proof of Theorem \ref{RHON}}

We now re-do the calculation but make the scaling $t =
\frac{A}{2N}$. We thus have
\begin{equation} \label{INT1} \begin{array}{lll} F_{SU(N)} (a, e^k;  A) & = &
 \frac{1}{2} \int_{S^1}  \log  H_{SU(2)} (\frac{A}{2N}, e^{i x}, e^{i \pi/2} ) d
\sigma_N * \overline{d \sigma_N}(e^{i x})  \\ & & \\ & &  -
\frac{1}{N} \log H_{SU(2)}(\frac{A}{2N}, 1, e^{i \pi/2})
\end{array} \end{equation}
We now obtain the limit by using the uniform off-diagonal
asymptotics of the central heat kernel.

The central heat kernel is related to the actual heat kernel by
$$\log \;H_{SU(2)}( t, e^{i x}, a) = \log  \int_{SU(2)}k_{SU(2)}  (t, e^{i x}, g^{-1} a g)
d g$$ where $k_{SU(2)}(t, x, y)$ is the heat kernel. Therefore, we
are interested in the uniform asymptotics of
\begin{equation} \log \;H_{SU(2)}( \frac{A}{2N} , e^{i x}, a) = \log  \int_{SU(2)}k_{SU(2)}
 (\frac{A}{2N}, e^{i x}, g^{-1} a g)
d g \end{equation} in $x$ for each $A$.   There exists a uniform
heat kernel parametrix for $k_{SU(2)}$ given by:
\begin{equation} k_{SU(2)}(t, u, v) \sim t^{- 3/2}
 e^{- \frac{d( u,v)^2}{ \pi t}} V(t, u, v) \end{equation}
 where $V(t, u,v)\sim  \sum_{j = 0}^{\infty} V_j(u, v) t^j$. The amplitude $V$ is to leading
 order the volume density in normal coordinates. We will never be evaluating it at a pair of conjugate
 points, so it is uniformly bounded below by a positive constant and introduces only lower order
 terms into the logarithm. See
 for instance \cite{Ka} for general results of this kind. Instead
 of a parametrix, the reader might prefer to use the
(\ref{HSU2}) in terms of Jacobi's theta function.  This formula
and also (\ref{HETA}) can be used to check the various constants
in the formula.

Thus, we have
\begin{equation}\label{INT2}  \log \;H_{SU(2)}( \frac{A}{2N} , e^{i x}, a) \sim  \log  \{(\frac{N}{A})^{3/2}
\int_{SU(2)} e^{- \frac{2 N}{ \pi A} d( e^{i x}, g^{-1} a g)^2}
V(A/2N, e^{i x}, g^{-1} a g) dg \},
\end{equation}
where $ V(A/2N, e^{i x}, g^{-1} a g)$ is a semiclassical amplitude
in $N$. The asymptotics are determined by the minimum point of the
phase $d( e^{i x}, g^{-1} a g)^2$, namely by the distance $d(e^{i
x}, C_a)$ from $e^{i x}$ to the conjugacy class of $a$. We note
that the conjugacy class $C(a) = \{g^{-1} a g: g \in SU(2)\}$ is a
 great (equatorial) $2$-sphere of radius $\pi/2$ from (the north pole)
 $I$.

There is a somewhat different expansion accordingly as $e^{i x} =
\pm 1$, $e^{i x} \in C(a)$ or for $e^{i x}$ not of this form,
which we will call a general element.  If $e^{i x}$ is a general
element, then there is a unique closest point $g^{-1} a g$ and the
phase is non-degenerate, so we obtain
\begin{equation} \label{ASYMP} \begin{array}{l} \int_{SU(2)} e^{- \frac{2 N}{\pi A} d( e^{i x},
g^{-1} a g)^2} V(A/2N, x, g^{-1} a g) dg  \\ \\ \sim  A^{-3/2} \;
e^{- \frac{2 N}{\pi A} d( e^{i x}, C_a)^2} \cdot
\frac{1}{\sqrt{\det Hess ( A^{-1} \;  d( e^{i x}, g^{-1} a
g)^2)}}.\end{array}
\end{equation}
We note that the powers of $A$ cancel.  If $e^{i x} \in C(a)$,
then since $e^{i x}$ also represents a point in the maximal torus,
$e^{i x} = a$ or $e^{i x} = a^{-1}$, hence the unique point of
minimal distance to $e^{i x}$ in $C(a)$ is of course $e^{i x}$
itself. There is no essential change in the calculation except
that the exponent vanishes. We now consider the behavior of
$\frac{1}{\sqrt{\det \; Hess (d( e^{i x}, g^{-1} a g)^2)}}$ as
$e^{i x} \to \pm 1$. In the case $e^{i x} \to 1$), for instance,
$\det Hess (A^{-1} \; d( e^{i x}, g^{-1} a g)^2) \sim \; |x|^3$
and hence
$$\log \det Hess (d( e^{i x}, g^{-1} a g)^2) \sim \log |x|\;\; \mbox{as}\;\; x \to 0.$$
At  $e^{i x} = \pm 1$, the
 entire $C(a)$ becomes a critical manifold for the phase.
 Exactly at the poles, we have
\begin{equation} H_{SU(2)}( t, \pm 1, a)  =  k_{SU(2)}  (\frac{A}{2N}, \pm 1,
a) \sim (\frac{N}{A})^{3/2} e^{ - 2 N \pi^2/ 4 \pi A}.
\end{equation} Thus, as $e^{i x} \to 1$,
\begin{equation} \label{ASYMP2} H_{SU(2)}(\frac{A}{2N}, e^{ix}, a)    \sim
\left\{ \begin{array}{ll}  |x|^{-3/2}   e^{- \frac{2 N}{\pi A} d(
e^{i x}, C_a)^2} , & x \not= 0 \\ & \\ (\frac{N}{A})^{3/2} e^{ - 2
N \pi^2/ 4 \pi A}, &  x = 0
 .\end{array} \right.
\end{equation}
Similarly at $x = \pi.$  We note that the exponent is continuous
and only the power of $N$ changes at the special points $x = 0,
\pi.$

Since we are interested in log asymptotics,  the factor
 $ e^{-
\frac{2 N}{\pi A} d( e^{i x}, C_a)^2}$ is dominant  as long as
the remainder   can be integrated against $d\sigma_N *
d\bar{\sigma}_N$. If this measure has a point mass at either $0$
or $\pi$, there are singularities at $x = 0, \pi$ in the
coefficients of the asymptotics which are not integrable $d \sigma
* d\bar{\sigma}$. We now discuss this point in detail.

We fix a positive constant $C > 0$ and  break up the integral
(\ref{INT1})  as
\begin{equation} \label{BREAKUP}  \begin{array}{l}   \int_{S^1
\backslash [-C/N, C/N] \cup [\pi -C/N, \pi + C/N] }  \log
H_{SU(2)} (\frac{A}{2N}, e^{i x}, e^{i \pi/2} ) d \sigma_N *
\overline{d \sigma_N}(e^{i x}) \\  \\  +
 \int_{ [-C/N, C/N] \cup [\pi -C/N, \pi + C/N]}\;  \log  H_{SU(2)} (\frac{A}{2N}, e^{i x}, e^{i \pi/2} ) d
\sigma_N * \overline{d \sigma_N}(e^{i x}). \end{array}
\end{equation}

It follows from the pointwise asymptotics in (\ref{ASYMP2}) that
\begin{equation} \label{INT1} \begin{array}{l} \int_{S^1
\backslash [-C/N, C/N] \cup [\pi -C/N, \pi + C/N] }  \log
H_{SU(2)} (\frac{A}{2N}, e^{i x}, e^{i \pi/2} ) d \sigma_N *
\overline{d \sigma_N}(e^{i x})  \\ \\ \sim   - \frac{2 N}{\pi A}
\; \int_{S^1 \backslash [-C/N, C/N] \cup [\pi -C/N, \pi + C/N] }
d(C(a), e^{ix})^2  d \sigma_N
* \overline{d \sigma}_N(e^{i x}). \end{array} \end{equation}

For the second integral of (\ref{BREAKUP}) over  $[-\frac{C}{N},
\frac{C}{N}]$ and $[\pi -C/N, \pi + C/N]$,  we  use the scaling
asymptotics of the heat kernel rather than its pointwise
asymptotics. Since the details are similar for both intervals we
only carry them out around the interval $[-\frac{C}{N},
\frac{C}{N}]$. Namely, we write $x = \frac{u}{N}$ with $|u| \leq
C$.  We have:
$$d( e^{i \frac{u}{N}}, g^{-1}
a g)^2  = (\frac{\pi}{2})^2 + \frac{u}{N} Q(u, N, A, g^{-1} a g)
$$ for a bounded smooth function $Q$, hence
\begin{equation} \label{SCALING} \begin{array}{lll}  \log \;H_{SU(2)}( \frac{A}{2N} , e^{i \frac{u}{N} }, a)
& \sim &
 \{(\frac{N}{A})^{3/2} e^{- \frac{2 N}{\pi A} \frac{\pi^2}{4}} \int_{SU(2)} e^{- Q(u, A, N)  g^{-1}
a g)^2} V(A/2N, e^{i \frac{u}{N}}, g^{-1} a g) dg \} \\ & & \\
& \sim & (\frac{N}{A})^{3/2} e^{- \frac{2 N}{A} \frac{\pi^2}{4}}[
q(u, A) + \frac{1}{N} q_1(u, A) + \cdots)\end{array}
\end{equation}
where $q = e^{- Q(u, A, 0})$ is strictly positive. Substituting
(\ref{SCALING}) into the second term of (\ref{BREAKUP}), we get
\begin{equation} \label{OINT} [- \frac{2 N}{\pi A} \frac{\pi^2}{4} - \frac{3}{2} \log (\frac{N}{A})]
 \int_{-C}^C d \sigma_N * d
\bar{\sigma}_N(\frac{u}{N})    - \int_{-C}^C Q (u, A) d \sigma_N *
d \bar{\sigma}_N(\frac{u}{N}).
\end{equation} Since $Q$ is bounded and $d \sigma_N(\frac{u}{N}) *
d \bar{\sigma}_N(\frac{u}{N})$ has mass at most $1$, the last term
is $O(1)$ as $N \to \infty$. Thus, the first term of (\ref{OINT})
dominates. Thus, we have
\begin{equation} \label{BREAKUP2}  \begin{array}{l}
 \int_{ [-C/N, C/N] \cup [\pi -C/N, \pi + C/N]}\;   \log  H_{SU(2)} (\frac{A}{2N}, e^{i x}, e^{i \pi/2} ) d
\sigma_N * \overline{d \sigma_N}(e^{i x})\\ \\
 \sim [- \frac{2 N}{\pi A} \frac{\pi^2}{4} - \frac{3}{2} \log (\frac{N}{A})]
 \int_{-C/N}^{/N} d \sigma_N * d
\bar{\sigma}_N.  \end{array} \end{equation}

We note that $d(C_a, 1) = d(C_a, -1) = \frac{\pi}{2}$. Since
$d(C_a, e^{ix})$ is continuous, we may rewrite
(\ref{OINT})-(\ref{BREAKUP2}) to leading order as
\begin{equation} \label{OINT2} - \frac{2 N}{\pi A} \;
 \int_{-C/N}^{C/N} d(e^{ix}, C_a)^2 \;\;d \sigma_N * d
\bar{\sigma}_N. \end{equation} We add this back to (\ref{INT1}) to
obtain the leading order term
\begin{equation}  - \frac{2 N}{\pi A} \;
\int_{S^1} d(C(a), e^{ix})^2  d \sigma_N
* \overline{d \sigma}_N(e^{i x}) \end{equation}
plus the canonical terms  $\frac{1}{N}  \log H_{SU(2)} (A/2N , 1,
a )$ and $\log \eta(\frac{iA}{2N})$ which are independent of
$d\sigma$. We then recognize that $d(e^{ix}, C_a) = \min
\{d(e^{ix}, e^{\pm i \pi/2})\},$ completing  the proof.

\qed

\begin{rem}

\begin{itemize}

\item It would be interesting to know when the leading order term
cancels, but even this cancellation would not cure the anomoly,
since there is no control over the rate of the  weak convergence
$d\sigma_N \to d\sigma$. Hence, there is no well-defined growth
rate of the lower order terms.

\item As noted above, the exponent in (\ref{ASYMP2}) is
continuous. Since we are taking the logarithm, only the exponent
is important to leading order, and that is why we obtain a unified
formula, even when $d\sigma$ has a point mass at $e^{ix} = \pm 1.$

\end{itemize}

\end{rem}

\section{\label{FR} Final Remarks}

Some final comments and remarks.

\begin{itemize}

\item It would of course be desirable to remove the assumption
that one of the matrices $U_1$ should be a prinicpal element of
type $\rho$. In \cite{F2}, H. Fegan states a product formula for
the full fundamental solution, but unfortunately this formula is
erroneous. It would be interesting to see if there exists some
form of MacDonald's identities for the full central heat kernel.

\item It would also be desirable to extend the results of this paper
to more general groups such as $U(N), SO(N), Sp(N)$. We specialize
to $G_N = SU(N)$  because it amply illustrates the main point of
this paper. Further,  MacDonald's identities pertain to simply
connected compact semi-simple groups, and presumably require some
modifications for $U(N), SO(N)$ etc.

\item
If we ignore the unexpected growth in $N$,  the phase transition
conjecture suggests that the limit free energy should equal
$$\begin{array}{l}  S(d\theta, d\sigma| A)  -
\frac{1}{2} \int_{S^1} \int_{S^1} \log |\sin \frac{\theta -
\phi}{2}| d\sigma(\theta) d\sigma( \phi)   - \frac{1}{2}
\int_{S^1} \int_{S^1}  \log |\sin \frac{\theta - \phi}{2}| d\theta
d \phi, \end{array}$$ The last two terms can be obtained from our
limit formula by unravelling the denominator in the logarithm.
Thus, aside from orders of magnitude, the phase transition
conjecture appears to state in these cases that
$$ S(d\theta, d\sigma| A)  =  \frac{1}{2 A} \int_{S^1}
d(C(a), e^{ix})^2  d \sigma
* \overline{d \sigma}(e^{i x}). $$
We have not been able to compare our results directly with the
physics predictions by solving the complex Burghers equation.

\item In addition to considering `pointwise' asymptotics of the
partition function, one could consider weak asymptotics where one
averages in various ways over the variables $U_1, U_2$, considers
the variance of the integrals and so on.

\item As mentioned in the introduction, it is possible that
an anomaly also occurs in Matytsin's conjectured asymptotics
\cite{M} at   the special character values $\chi_R(a)$.

\end{itemize}

\end{document}